\documentclass{iaus}
\usepackage{graphicx}

\title[IAUS266.~~The origin of very wide binary systems] 
{The origin of very wide binary systems}

\author[M. B. N. Kouwenhoven et al.]
{M. B. N. Kouwenhoven,$^{1,2}$
 S. P. Goodwin,$^2$
 Richard J. Parker,$^2$\\
 M. B.~Davies,$^3$
 D. Malmberg$^3$
 \and P. Kroupa$^4$
}

\affiliation{
$^1$Kavli Institute for Astronomy and Astrophysics, Peking University, Yi He Yuan Lu 5,\\ Hai
Dian District, Beijing 100871, P.\,R. China \\ email: {\tt kouwenhoven@kiaa.pku.edu.cn} \\
[\affilskip]
$^2$University of Sheffield, Hicks Building, Hounsfield
Road, Sheffield S3\,7RH, UK \\
[\affilskip]
$^3$Lund Observatory, Box\,43, SE-221\,00, Lund, Sweden \\
[\affilskip]
$^4$Argelander Institute for Astronomy, University of
Bonn, Auf dem H\"{u}gel 71, 53121, Bonn, Germany\\
}

\pubyear{2010}
\volume{266}  
\pagerange{1--4}
\jname{Star clusters: basic galactic building blocks}
\editors{R. de Grijs \& J. R. D. L\'epine, eds.}
\begin{document}

\maketitle

\begin{abstract}
The majority of stars in the Galactic field and halo are part of
binary or multiple systems. A significant fraction of these systems
have orbital separations in excess of thousands of astronomical units,
and systems wider than a parsec have been identified in the Galactic
halo. These binary systems cannot have formed through the `normal'
star-formation process, nor by capture processes in the Galactic
field. We propose that these wide systems were formed during the
dissolution phase of young star clusters. We test this hypothesis
using $N$-body simulations of evolving star clusters and find wide
binary fractions of $1-30$\%, depending on initial
conditions. Moreover, given that most stars form as part of a binary
system, our theory predicts that a large fraction of the known wide
`binaries' are, in fact, multiple systems.  
\end{abstract}

\firstsection 

\section{Observations and formation of very wide binary systems}

Numerous wide ($10^3$~AU$-0.1$~pc) binary systems are known to exist
in the Galactic field and halo (e.g., Close et al. 1990; Chanam{\'e}
\& Gould 2004). These binaries are usually detected as
common-proper-motion pairs (e.g., Wasserman \& Weinberg 1991;
L{\'e}pine \& Bongiorno 2007) or through statistical methods (e.g.,
Bahcall \& Soneira 1981; Garnavich 1988). Approximately 15\% of the
wide binaries in the field have a semi-major axis in the range
$10^3~{\rm AU} < a < 0.1$~pc (Duquennoy \& Mayor 1991; see also Poveda
et al. 2007). The origin of these wide binaries cannot be explained by
star formation or by dynamical interactions in the Galactic field, and
has long been a mystery.

The majority of stars are known to form in embedded clusters (Lada \&
Lada 2003). Wide binary systems cannot have formed through clustered
star formation, simply because their semi-major axis is comparable to
the size of a typical embedded cluster. Even if it were possible to
form wide binaries in star clusters, they would immediately be
destroyed by dynamical interactions. Although most stars are believed
to have formed in star clusters, a fraction ($< 10-30$\%) may form
through diffuse (isolated) star formation. However, this fraction
makes it difficult to account for the $\sim 15$\% wide binaries in the
Galactic field. If wide binaries are formed through diffuse star
formation, then the majority of stars formed in this process must form
a wide binary, implying a star-formation process that is fundamentally
different from that of clustered star formation.

Binary systems of any separation may also form through dynamical
capture processes. To form a binary system from two interacting stars,
a third star should be present to serve as energy sink. Each of the
three stars should be at the same location at the same time, and
should have precisely those velocities and impact angles that allow
the formation of a new binary system. These conditions are extremely
rare in the Galactic field, resulting only in a handful of binaries
over the lifetime of the Galaxy (Goodman \& Hut 1993). Note that
binaries can form through the capture process in the cores of star
clusters through this process. However, wide binary systems cannot
form in these environments, simply because they have a semi-major axis
comparable to (or even larger than) star cluster cores.

We propose that wide binaries are formed during the dissolution of
young star clusters (Kouwenhoven et al., in prep). Young clusters
generally have a short lifetime and dissolve into the field stellar
population within typically 20~Myr (e.g., Mengel et al. 2005; Bastian
et al. 2005; de Grijs \& Parmentier 2007). During this dissolution
process, two stars may escape the cluster in the same direction, with
similar velocities. Although initially unbound, these two stars may
form a new binary system after having left the gravitational potential
of the cluster. In this process, the separation between the two stars
is typically on the order of the size of their natal cluster at the
time of dissolution, i.e., of order a parsec.

\section{$N$-body simulations and results}

\begin{figure}[tb]
  \begin{center}
    \includegraphics[width=\textwidth]{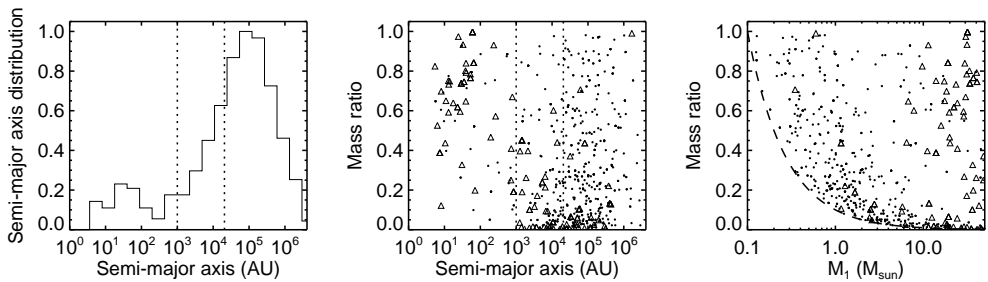}
    \caption{The binary population resulting from a Plummer model with
$N=1000$ and a virial radius of $0.1$~pc (50 runs), showing the
semi-major-axis distribution {\it (left)}, the correlation between
mass ratio $q$ and semi-major axis $a$ {\it (middle)} and between
primary mass and mass ratio {\rm (right)}. Binary and multiple systems
are indicated with circles and triangles, respectively. The vertical
dashed lines indicate $a = 10^3$~AU and $a = 0.1$~pc, respectively,
the semi-major-axis range we consider in our analysis. The dashed
curve in the right-hand panel indicates the minimum mass ratio $q_{\rm
min}(M_1) = M_{\rm min}/M_1$ (Kouwenhoven et al. 2009).  }
  \end{center}
\end{figure}

It is theoretically possible to form a wide binary in the star cluster
dissolution phase and it is important to find out whether this process
happens frequently enough to account for the observed wide binary
population. To this end, we carry out $N$-body simulations of evolving
star clusters using the {\sc starlab} package (Portegies Zwart et
al. 2001). We randomly draw $N$ single stars from the Kroupa (2002)
initial mass function in the mass range $0.1-50$ M$_\odot$. The stars
are given positions and velocities according to two models, (i) the
spherical and homogeneous Plummer (1911) model and (ii) a fractal
stellar distribution with fractal dimension 1.5, following the
prescriptions of Goodwin \& Whitworth (2004). The simulations are run
until the star cluster is completely dissolved. We only consider the
wide binary population in the semi-major-axis range $10^3~{\rm AU} < a
< 0.1$~pc, where the upper limit roughly corresponds to the stability
limit in the Galactic field. We find the dependence on initial
conditions by varying the cluster mass, size and morphology. Our main
results are the following.

\begin{enumerate}
\item The resulting wide binary fraction is $1-30$\%, depending on the
initial properties of the star cluster. The wide binary fraction
increases strongly with increasing initial substructure of the
cluster, and decreases with increasing cluster mass.

\item The dissolution of star clusters results in a binary population
with a bimodal semi-major-axis distribution. At small separations, we
find a {\em dynamical peak}, consisting of binaries formed dynamically
in the centre of the star cluster. At large separations, we find a
{\em dissolution peak}, which contains the binary systems formed
during the dissolution process of the star cluster. As an example, we
show in Figure~1 the binary population resulting from a star cluster
with $N=10^3$ single stars and an initial radius of 0.1~pc. The
resulting semi-major axis scales linearly with the initial size of the
star cluster.

\item Wide-binary formation is a process where two random stars
encounter each other. When two stars are nearby in phase space, their
mutual gravitational attraction (i.e., their combined mass) determines
whether they form a new binary. The resulting mass-ratio distribution
can therefore be characterised by gravitationally focused random
pairing (e.g., Gaburov et al. 2008). The resulting eccentricity
distribution is thermal (e.g., Heggie 1975) and there is no
correlation between the orbital orientation and the stellar rotation
vectors.
\end{enumerate}

\section{Implications for higher-order multiplicity}

In the previous section we described simulations of star clusters that
initially consist of single stars only. In reality, however, most
stars (possibly even all stars) are formed as part of binary or
multiple systems (e.g., Duquennoy \& Mayor 1991; Fischer \& Marcy
1992; Kroupa 1995; Kouwenhoven et al. 2005, 2007; Zinnecker \& Yorke
2007). To first order, primordial binaries with separations much
smaller than those of wide binaries can be dynamically approximated as
single stars. Therefore, for a star cluster with primordial binaries
we expect to find approximately the same wide-binary population as
described above. But in this case, the wide `binaries' are, in fact,
wide triple or quadruple systems, where one or both of the subsystems
consists of a primordial binary. Our $N$-body simulations of clusters
with primordial binaries show that this is indeed the case.  The ratio
between the newly formed wide binaries, triples and quadruples, is a
function of the primordial binary fraction. For example, for a
primordial binary fraction of 50\%, we expect a
binary:triple:quadruple ratio of 1:2:1. By accurately measuring the
multiplicity of wide-binary systems, we can therefore constrain the
primordial binary fraction.

\section{Summary}

A significant fraction of the stars in the Galactic field and halo are
part of wide ($a>10^3$~AU) binary systems. The origin of these binary
systems cannot be explained by clustered star formation, nor by
dynamical capture in the Galactic field and halo.  We propose that
these binaries are formed during the dissolution process of young star
clusters, where two initially unbound stars escape in a similar
direction and form a bound wide binary (Kouwenhoven et al., in prep).
We test this hypothesis using $N$-body simulations and find that we
are able to reproduce wide-binary fractions of $1-30$\% (in the
semi-major-axis range $10^3~{\rm AU} < a < 0.1$~pc), depending on the
initial conditions.  Moreover, as most stars form as part of a binary
system, we predict that a large fraction of the known wide `binaries'
are, in fact, triple or quadruple systems, and the ratio of wide
binary, triple and quadruple systems therefore provides an indication
of the primordial binary fraction.
\newpage
\begin{acknowledgements}
MBNK was supported by the Peter and Patricia Gruber Foundation through
a PPGF fellowship, the Kavli Foundation, the Peking University One
Hundred Talent Fund (985) and by STFC (grant PP/D002036/1). RJP
acknowledges support from STFC.
\end{acknowledgements}

\end{document}